\documentclass[prl,superscriptaddress,twocolumn]{revtex4}
\usepackage{amsmath}
\usepackage{graphicx}
\usepackage{dcolumn}
\usepackage{bm}
\begin{document}
\title{Interaction and filling induced quantum phases of \\dual Mott insulators of bosons and fermions}
\author{Seiji Sugawa*}
\affiliation{Department of Physics, Graduate School of Science,
Kyoto University, Kyoto 606-8502, Japan}
\email{ssugawa@scphys.kyoto-u.ac.jp}
\author{Kensuke Inaba}
\affiliation{NTT Basic Research Laboratories, NTT Corporation, Atsugi 243-0198, Japan}
\affiliation{JST, CREST, Chiyoda-ku, Tokyo 102-0075, Japan}
\author{Shintaro Taie}
\affiliation{Department of Physics, Graduate School of Science, 
Kyoto University, Kyoto 606-8502, Japan}
\author{Rekishu Yamazaki}
\affiliation{Department of Physics, Graduate School of Science, 
Kyoto University, Kyoto 606-8502, Japan}
\affiliation{JST, CREST, Chiyoda-ku, Tokyo 102-0075, Japan}
\author{Makoto Yamashita}
\affiliation{NTT Basic Research Laboratories, NTT Corporation, Atsugi 243-0198, Japan}
\affiliation{JST, CREST, Chiyoda-ku, Tokyo 102-0075, Japan}
\author{Yoshiro Takahashi}
\affiliation{Department of Physics, Graduate School of Science, 
Kyoto University, Kyoto 606-8502, Japan}
\affiliation{JST, CREST, Chiyoda-ku, Tokyo 102-0075, Japan}
\date{\today}
\begin{abstract}
Many-body effects are at the very heart of diverse phenomena found in condensed-matter physics.
One striking example is the Mott insulator phase where conductivity is suppressed as a result of 
a strong repulsive interaction.
Advances in cold atom physics have led to the realization of the Mott insulating phases of atoms in an optical lattice, 
mimicking the corresponding condensed matter systems.
Here, we explore an exotic strongly-correlated system of {\it Interacting Dual Mott Insulators} of bosons and fermions.
We reveal that an inter-species interaction between bosons and fermions drastically modifies each Mott insulator, 
causing effects that include melting, generation of composite particles, an anti-correlated phase, and complete phase-separation. 
Comparisons between the experimental results and numerical simulations indicate intrinsic adiabatic heating and cooling 
for the attractively and repulsively interacting dual Mott Insulators, respectively.
\end{abstract}
\maketitle
\parskip=0pt
The Mott insulator transition is a dramatic manifestation of quantum many-body effects. 
In condensed matter systems, strong repulsive interactions between electrons lead to the insulating behavior of materials,
when the commensurate condition for the density is satisfied. 
Several interesting phenomena also have been discovered around the Mott transition point,
including diverse phases showing anomalous fluctuations and orderings in the spin, charge, and orbital degrees of freedom. 
These are thought to be deeply connected with the mechanism of high-temperature superconductivity in cuprates and colossal magnetoresistance in manganites\cite{ImadaReview}.  
\par
Recently, Mott insulating phases have been studied by ultracold atoms confined within an optical lattice \cite{JakschReview, BlochReview}
where three mutually perpendicular standing laser waves create a periodic potential.
Making use of the high degree of control possible with  atomic systems, both the bosonic superfluid-Mott insulator transition \cite{Fisher1989,Jaksch1998,Greiner2002,Spielman2007} 
and the fermionic metal-Mott insulator transitions\cite{Fermi_Mott_ETH, Fermi_Mott_MPQ} have been experimentally demonstrated.
The detailed behavior of the Mott insulating phase has been well studied by various spectroscopic techniques
\cite{BlochReview, Campbell2006a, Folling2006, Chin_2DMott, Bakr2010, Sherson2010}. 
\par
Given the important role of the interaction effects in these systems, one could expect drastic changes when two different Mott insulators with strong interspecies interaction are combined together in the same lattice.
Such a system of strongly interacting dual Mott insulator shown in Fig. 1 offers a unique possibility for exploring rich phases that are not accessible with
a single Mott insulator system.
In particular, in addition to interspecies interactions, the relative fillings of two species 
bring a different kind of diversity to this system, which provides us with a novel paradigm such as {\it filling induced phases}\cite{ImadaReview}. 
\par
Theoretical work in this area has been rich. Various novel phases have been theoretically discussed for mixtures of bosons and polarized fermions\cite{Titvinidze2008, Lewenstein2004}, 
Bose-Bose mixtures\cite{Hubener2009, Capogrosso-Sansone2010}, and Fermi-Fermi mixtures\cite{Iskin2007}.
Pioneering experimental work has studied the Bose-Fermi mixture of 
rubidium ($^{87}$Rb) and spin polarized potassium ($^{40}$K),
where impurity fermions change the bosonic superfluid-Mott insulator transition\cite{BF_ETH, BF_Hamburg, BF_MPQ}.
However, until now, the study of interaction and filling induced phases in the dual Mott insulating regime was not explored.
This is also the case for Bose-Bose mixtures \cite{BB_LENS, BB_MIT,BB_Stony} and Fermi-Fermi mixture \cite{FermiFermi,Taie2010}.
\par
In this article, we investigate strongly-correlated systems of attractively and repulsively interacting dual Mott insulators of bosonic and fermionic atoms.
To characterize the dual Mott insulators, we measure double occupancy of bosonic and fermionic atoms 
and pair occupancy of boson and fermion for varying numbers of atoms.
From these measurements in combination with numerical calculation, 
we reveal how the boson-fermion interspecies interaction and the number of atoms drastically modify 
the basic behavior of each Mott insulator, both bosonic and fermionic.
With repulsive interactions, each Mott insulator is phase separated or turns into anti-correlated mixed Mott insulating phase depending on the number of atoms.
In contrast, we observe that in the attractively interacting dual Mott insulator
various composite particles of bosons and fermions are formed.
Comparison between the experimental results and the numerical simulation strongly indicates 
intrinsic adiabatic heating and cooling for the attractively and repulsively interacting dual Mott Insulators, respectively.

		\begin{center}
		\begin{figure*}[bht]
		\includegraphics[width=13cm]{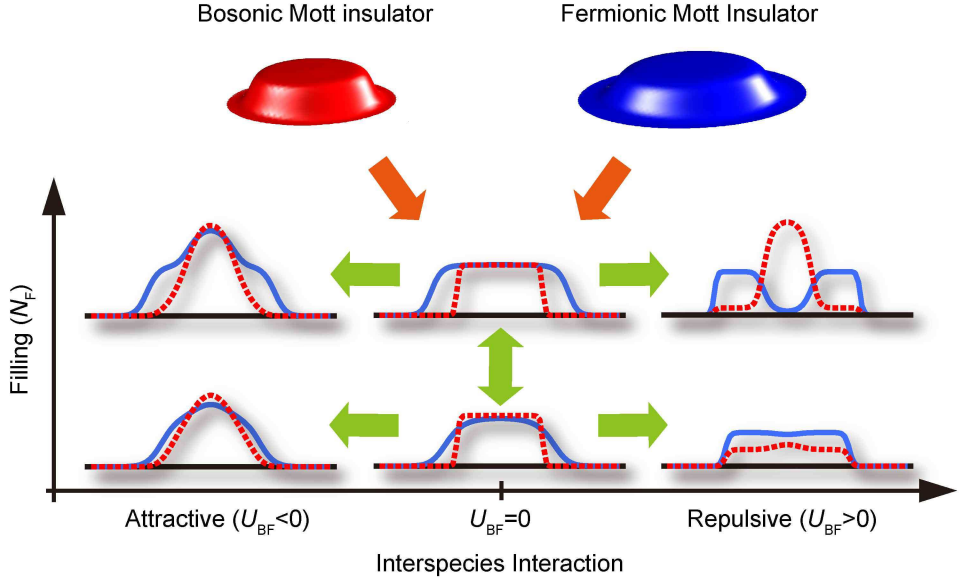}
		\caption{{\bf Schematic illustration of effects of interspecies interaction and filling on dual Mott insulators of bosons and fermions.} 
		Depending on the sign of the interspecies interaction and filling, 
		the density distributions of each Mott insulator of unit occupancy is dramatically altered. 
		The red dashed and blue lines represent the spatial distributions of bosons and fermions, respectively.}
		\end{figure*}
		\end{center}
\section{The system of dual Mott insulators of bosons and fermions} 
Our experiment is performed with a quantum degenerate gas of bosonic and fermionic isotopes of ytterbium(Yb) atoms subjected to a three-dimensional optical lattice of simple cubic symmetry with a lattice constant of $266$ nm. 
An attractively interacting Bose-Fermi dual Mott insulator is prepared with the bosonic isotope of  $^{170}$Yb mixed with the fermionic isotope of $^{173}$Yb.
Separately, a repulsively interacting dual Mott insulator is prepared with the bosonic isotope of  $^{174}$Yb mixed with $^{173}$Yb. 
In both types of dual Mott insulators, the fermionic isotope of $^{173}$Yb is prepared in an almost equal population of six nuclear spin components of $I=5/2$.
The relevant s-wave scattering lengths of Yb bosons and fermions used in this work are as follows:
$a_{170}$=3.38 nm, $a_{174}$=5.55 nm and $a_{173}$=10.55 nm
as well as the interspecies s-wave scattering lengths, $a_{170-173}$=$-$4.30 nm and $a_{174-173}$=7.34 nm.
Here, it is noted that $a_{173}$, $a_{170-173}$ and $a_{174-173}$ are independent of nuclear-spin components of $^{173}$Yb\cite{Kitagawa2PA}.
\par
The physics of a Mott insulator is well described by a single-band Hubbard model.
It considers particle-tunneling between adjacent lattice sites and on-site interaction. 
In optical lattice experiments, the underlying harmonic confinement gives rise to an additional spatially dependent energy offset. 
This has important influence on the spatial structure of the Mott insulators\cite{Batrouni2002}. 
\par
As a result, our system of dual Mott insulators of bosons and fermions
can be described by the Hubbard Hamiltonian of
bosons($\hat{H}_{\rm B}$) and fermions($\hat{H}_{\rm F}$) plus
boson-fermion on-site interactions;
\begin{align}
\hat{H}_{\rm BF} =& \hat{H}_{\rm B} + \hat{H}_{\rm F} + U_{\rm BF}
\sum_{i,\sigma} \hat{n}_{{\rm B},i} \hat{n}_{{\rm F},i,\sigma} ,\\
\hat{H}_{\rm B} =& -J \sum_{\langle i,j \rangle} \hat{b}_{i}^{\dagger}
\hat{b}_{j}
+ \frac{U_{\rm BB}}{2} \sum_{i} \hat{n}_{{\rm B},i} (\hat{n}_{{\rm B},i}-1) \\\notag
&+\sum_{i,\sigma} \epsilon_i \hat{n}_{{\rm B},i},\\
\hat{H}_{\rm F} =& -J \sum_{\langle i,j \rangle,\sigma }
\hat{f}_{i,\sigma}^{\dagger} \hat{f}_{j,\sigma}
+\frac{U_{\rm FF}}{2} \sum_{i,\sigma\neq\sigma^{'}} \hat{n}_{{\rm
F},i,\sigma} \hat{n}_{{\rm F},i,\sigma^{'}} \\\notag
&+\sum_{i,\sigma} \epsilon_i \hat{n}_{{\rm F},i,\sigma} ,
\end{align}
where $\langle i,j \rangle$ denotes summation over nearest-neighbors and
$\sigma \in \{1,2,...,6\}$ denotes the spin component of fermionic atoms.
$\hat{b}_{i}^{\dagger}$ ($\hat{f}_{i,\sigma}^{\dagger}$) and
$\hat{b}_{i}$ ($\hat{f}_{i,\sigma}$) are creation and annihilation
operators for bosons (fermions with $\sigma$ spin).
$\hat{n}_{{\rm B},i}= \hat{b}_{i}^{\dagger} \hat{b}_{i}$ ($\hat{n}_{{\rm
F},i,\sigma}= \hat{f}_{i,\sigma}^{\dagger} \hat{f}_{i,\sigma}$)
are the number operators for bosons (single spin component fermions).
$U_{\rm BB}$, $U_{\rm FF}$, and $U_{\rm BF}$ represent on-site
interaction energies between bosons, fermions with different spin
components, and boson-fermion pairs, respectively.
Here $U_{\rm BB} < \lvert U_{\rm BF} \rvert < U_{\rm FF}$ is satisfied
in both systems.
$J$ is the tunneling matrix element, and $\epsilon_{i}$ is the energy
offset due to an external confinement potential.

\section{Double and Boson-Fermion Pair occupancy measurements} 
In order to investigate the dual Mott insulators, 
we measure site occupancies, such as double occupancy of either bosons or fermions and pair occupancy of boson and fermion, which are
schematically depicted in Fig. 2(a).
The suppression of double occupancy is a key feature of the incompressibility of a Mott insulator\cite{Fermi_Mott_ETH}
and the pair occupancy of boson and fermion is a direct measure of their overlap.
\par
To measure the number of double occupancy, we use a one-color photoassociation (PA) technique.
For the site with two atoms, the irradiation of the PA laser beam results in the formation of excited state molecule which escapes from the trap,
which can be seen through atom loss.
We apply a strong enough PA laser beam so that all the doubly occupied sites are depleted with high efficiency.
As occupancy of more than two bosons or two fermions is negligible in the current experimental parameters, 
the atom loss corresponds to the absolute number of double occupancy\cite{Rom2004}.
Figure 2(b) shows the atom loss due to a strong PA laser beam where the number of remaining atoms is obtained from absorption images.
\par
An illustrating example of PA spectra is shown in Fig. 2(c) with bosonic homonuclear resonances in an optical lattice.
In the case of a pure bosonic cloud with a Mott plateau of one atom per site, there is essentially no atom loss, as expected.
When we add fermions to the bosonic Mott insulator, however, we observe a significant atom loss,
meaning sizable double occupancy of bosons.
This suggests that a fermion admixture induces a drastic change in the bosonic Mott insulating phase. 
This novel phenomenon of fermion-induced bosonic double occupancy is discussed in detail in the following sections.
\par
The pair occupancy of boson and fermion is measured in the same manner 
by utilizing heteronuclear PA resonances between bosonic and fermionic isotopes. 
The detailed information on the resonances is given in Methods.

		\begin{center}
		\begin{figure}[th]
		\includegraphics[width=8.5cm]{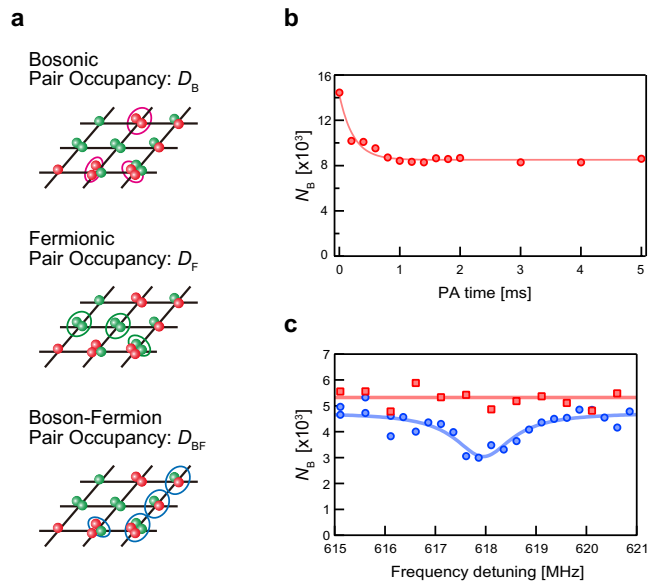}
		\caption{{\bf Photoassociation  and occupancy measurements in an optical lattice.}
		{\bf (a)} Schematic illustration of three types of occupancy in an optical lattice.
		A bosonic atom is represented by red, and fermionic by green.
		The number of atoms circled with a solid line is measured in the respective measurement.
		{\bf (b)} Decay of the number of bosonic atoms due to the irradiation of the PA laser in the bosonic Mott insulator. 
		Only the doubly occupied sites are depleted by the PA laser.
		{\bf (c)} PA resonance spectra of $^{170}$Yb bosons
		in an attractively interacting Bose-Fermi system($^{170}$Yb-$^{173}$Yb, blue circles) 
		and in a pure bosonic cloud that forms Mott plateau of unit filling($^{170}$Yb, red squares).  
		The baseline of the spectrum of the pure bosonic cloud is shifted vertically by 1$\times$10$^3$ for clarity.
		The solid blue line is a fit with a Lorentzian curve, and the solid red line represents no resonance. 
		}
		\end{figure}
		\end{center}

		\begin{center}
		\begin{figure*}[tbp]
		\includegraphics[width=16cm]{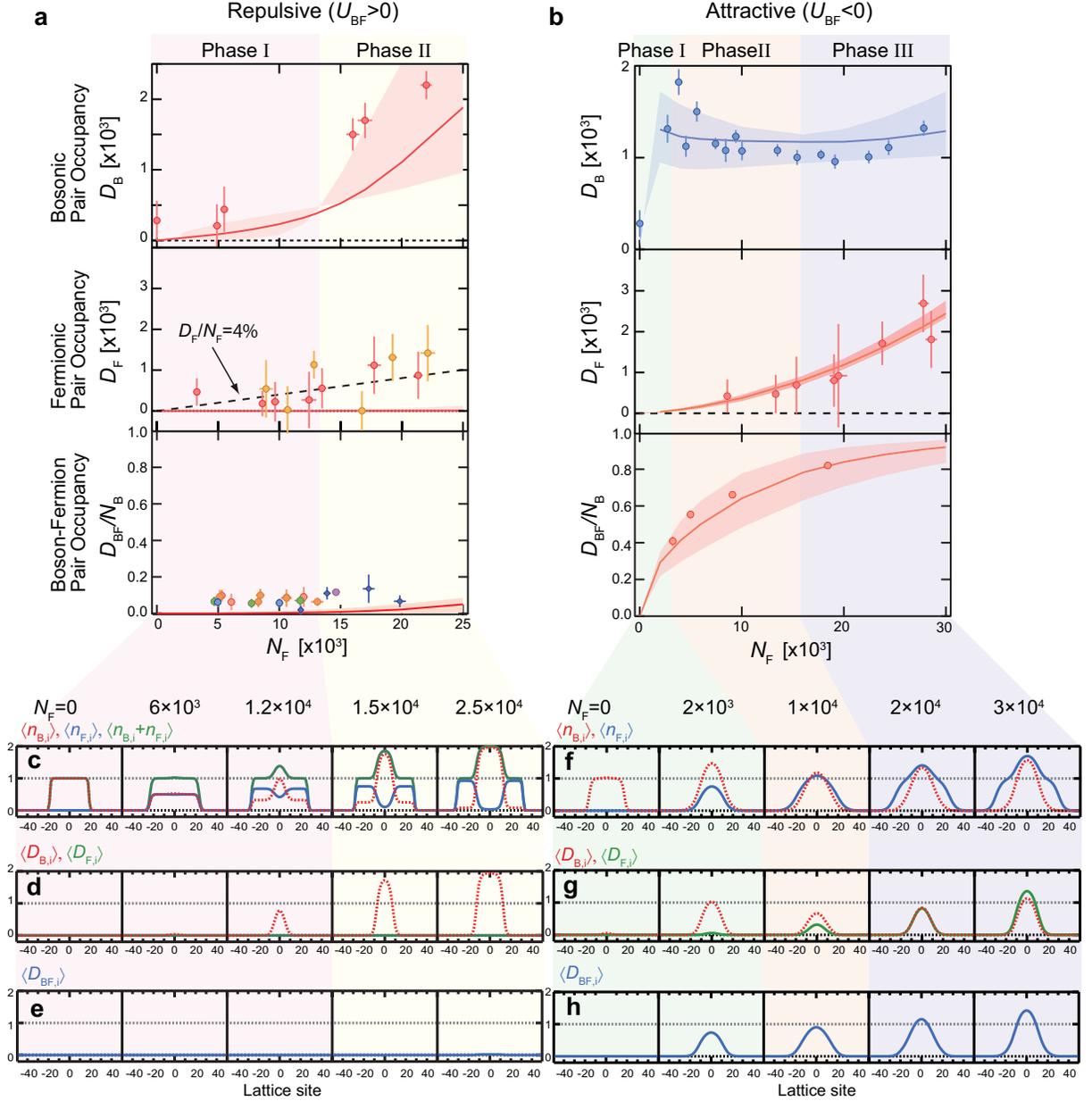}
		\caption{
		{\bf Occupancy measurements and numerical simulation of occupancy distributions.}
		{\bf (a)}Measured pair occupancies for the repulsively interacting system;
		{\bf (a)}(top) Pair occupancy of bosonic atoms ($D_{\rm B}$) for $N_{\rm B}=6\times 10^3$ represented by red circles, 
		{\bf (a)}(middle) Pair occupancy of fermionic atoms ($D_{\rm F}$) for $N_{\rm B}=9\times 10^3$ represented by red circles, and the
		fermionic pair occupancy for pure fermionic atoms (yellow circles) shown for comparison,
		{\bf (a)} (bottom) Boson-fermion pair occupancy ($D_{\rm BF}/N_{\rm B}$) for $N_{\rm B}=6\times 10^3$ (red circles), 
		$N_{\rm B}=8.5\times 10^3$ (orange circles), 
		$N_{\rm B}=13\times 10^3$ (green circles), $N_{\rm B}=15\times 10^3$ (blue circles), and $N_{\rm B}=20\times 10^3$ (purple circles).
		The data with almost the same mean trap frequency but with slightly different geometry	($2\pi \times$(150, 30, 150) Hz)	
		are represented by blue diamonds ($N_{\rm B}= 5.5\times 10^3$).
		Note that the pair occupancy is divided by the number of bosonic atoms to show fractional overlap between bosons and fermions.
		{\bf (b)} Measured pair occupancies for the attractively interacting system;
		{\bf (b)}(top) Pair occupancy of bosonic atoms for $N_{\rm B}=5\times 10^3$ represented by blue circles,
		{\bf (b)}(middle) Pair occupancy of fermionic atoms for $N_{\rm B}=5\times 10^3$ represented by red circles,
		{\bf (b)}(bottom) Boson-fermion pair occupancy divided by the number of bosonic atoms ($D_{\rm BF}/N_{\rm B}$) (red circles).
		The solid lines in {\bf (a)} and {\bf (b)} correspond to the numerical simulation for {\bf (a)} $T_{\rm{ini}}$=35 nK and {\bf (b)} $T_{\rm{ini}}$=45 nK 
		 where $T_{\rm{ini}}$ is the initial atom temperature before loading into an optical lattice. 
		The numerical results with different initial temperatures by $\pm$ 5 nK are also represented by the shaded areas along the solid lines.
		{\bf (c)-(e)}Numerical simulation for the repulsively interacting system with $N_{\rm B}=6\times 10^3$, $T_{\rm{ini}}$=30 nK;
		{\bf (c)}Column density distributions for bosons ($\langle n_{\rm B, i} \rangle$) (red dashed lines), 
		fermions ($\langle n_{\rm F, i} \rangle$)(blue lines),
		and their summation ($\langle \hat{n}_{\rm B, i} + \hat{n}_{\rm F, i} \rangle$)(green lines),
		{\bf (d)} On-site double occupancy distributions of bosons ($\langle D_{\rm B, i}\rangle$) (red dashed lines)
		and fermions ($\langle D_{\rm F, i}\rangle$)  (green lines),
		{\bf (e)} On-site boson-fermion pair occupancy distribution ($\langle D_{\rm BF, i}\rangle$).
		{\bf (f)-(h)}Numerical simulation for the attractively interacting system with $N_{\rm B}=5\times 10^3$, $T_{\rm{ini}}$=40 nK;		
		{\bf (f)} Column density distributions of bosons ($\langle n_{\rm B, i} \rangle$) (red dashed lines) and 
		fermions ($\langle n_{\rm F, i} \rangle$)(blue lines),
		{\bf (g)} On-site double occupancy distributions of bosons ($\langle D_{\rm B, i}\rangle$) (red dashed lines)
		 and fermions ($\langle D_{\rm F, i}\rangle$) (green lines),
		{\bf (h)} On-site boson-fermion pair occupancy distribution ($\langle D_{\rm BF, i}\rangle$).
		}
		\end{figure*}
		\end{center}
		
\section{Repulsively interacting dual Mott insulators}  
First, we discuss the case of repulsively interacting dual Mott insulators.
The results of the measurements are shown in Fig. 3 (a),
where the bosonic double occupancy ($D_{\rm B}$), the fermionic double occupancy ($D_{\rm F}$), and boson-fermion pair occupancy ($D_{\rm BF}$)
are plotted as a function of the number of fermionic atoms ($N_{\rm F}$).
The numerical calculations for density distributions for bosons ($\langle \hat{n}_{\rm B, i} \rangle$), fermions ($\langle \hat{n}_{\rm F, i} \rangle$)
and the bosons and fermions ($\langle \hat{n}_{\rm B, i} +\hat{n}_{\rm F, i} \rangle$) are shown
in Fig. 3 (c) for various numbers of fermionic atoms.
The calculated distributions of double occupancies for bosons ($\langle D_{\rm B, i}\rangle$) and fermions ($\langle D_{\rm F, i}\rangle$) are shown in Fig. 3 (d),
and those of boson-fermion pair occupancy ($\langle D_{\rm BF, i}\rangle$) in Fig. 3(e).

From these results we have found that the system is characterized by the two distinct phases depending on $N_F$.
Here we begin by explaining the calculated results.
As shown in Fig. 3(c), when $N_F<10^4$,
the density distributions of bosons ($\langle n_{\rm B, i} \rangle$) and fermions ($\langle \hat{n}_{\rm F, i} \rangle$) apparently overlap with each other, 
 while the total density distribution ($\langle \hat{n}_{\rm B, i} +\hat{n}_{\rm F, i} \rangle$) clearly satisfies $n_{{\rm B},i},$+$n_{{\rm F},i}$=1,
 meaning the formation of the Mott plateau at the unit filling.
In addition, from Fig. 3(d) and (e), bosonic, fermionic and boson-fermion pair occupancies take values near zero. 
We thus identify this phase as a novel kind of mixed Mott state of bosons and fermions by noting that a total set of bosons and fermions satisfies comensurability, but species separately does not.
\par
As the number of fermionic atoms increases, a different phase appears.
Figure 3 (c) clearly shows that bosons and fermions become phase-separated for larger numbers of fermions. 
The boson-fermion pair occupancy also takes a value near zero, even at larger numbers of fermions, as shown in Fig. 3(e).
This phase-separation can be understood from the relative strength of the on-site interaction parameters ($U_{\rm BB} < U_{\rm BF}  <  U_{\rm FF}$).
Since the fermion on-site interaction $U_{\rm FF}$ is the largest, it is energetically favorable for fermions to spatially extend to avoid double occupancy.
Due to the Bose-Fermi repulsive interaction $U_{\rm BF}$, bosons and fermions repel each other, which results in a phase-separation.
Figure 3(c) also shows that bosonic double occupancy grows in the center of the trap.
\par
The experimental results of the occupancy measurements (Fig. 3(a)) confirm the above-mentioned features of numerical simulation, 
demonstrating that these phases are realized in our experiments.
First, the bottom panel of Fig. 3(a) shows strong suppression of $D_{\rm BF}$, lower than 10 $\%$ of $N_{\rm B}$,
which is consistent with the numerical simulation (Fig. 3(e)).
$D_{\rm F}$ stays small, as shown in the middle panel, which is consistent with $\langle D_{\rm F, i} \rangle$ close to zero(see Fig. 3 (d)).
The top panel shows the suppression of $D_{\rm B}$ in the mixed Mott regime at small numbers of fermions, 
which is consistent with the numerical calculation of $\langle D_{\rm B, i} \rangle$ in Fig. 3 (d).
At larger numbers of fermions, we observe a large $D_{\rm B}$ in Fig. 3 (a). 
We can explain this phenomenon by the strong compression exerted on bosons from the surrounding fermion shell,
due to the strong interspecies repulsion.
Numerical calculations using the temperature as the single fitting parameter quantitatively explain all the experimental data.

		\begin{center}
		\begin{figure}[htbp]
		\includegraphics[width=6cm]{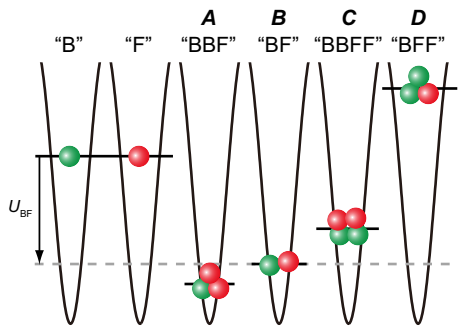}
		\caption{
		{\bf Various types of composite particles in the attractively interacting system.}
		The on-site interaction energy of the composite particles depends on the composition of particles.
		Type A, B, C, and D composite particle have energies of $U_{\rm BB} + 2 U_{\rm BF}=1.19 U_{\rm BF}$,
		$U_{\rm BF}$, $U_{\rm BB} + U_{\rm FF} + 4 U_{\rm BF}=0.67 U_{\rm BF}$, 
		and $U_{\rm FF} + U_{\rm BF}=- 0.63 U_{\rm BF}$, respectively.
		}
		\end{figure}
		\end{center}

\section{Attractively interacting dual Mott insulators}   
A remarkable difference between the repulsively and the attractively interacting system is observed. This highlights the crucial role of interspecies interaction in strongly interacting dual Mott insulators of bosons and fermions. 
\par
The experimental results are summarized in Fig. 3 (b). The clear departure from the behavior of the repulsively interacting system is seen in the boson-fermion pair occupancy measurement, where we observe the significantly large pair occupancy. The behavior of bosonic double occupancy, shown in the top panel, is also quite different. The measured $D_{B}$ shows a characteristic nonmonotonic dependence on $N_{\rm F}$ values and three phases can be identified. 

In phase I, where $N_{\rm F}$ is smaller than $2.5 \times 10^3$,
$D_{\rm B}$ rapidly increases as $N_{\rm F}$ increases. 
In the regime of intermediate number of fermionic atoms (phase II),
$D_{\rm B}$ gradually decreases, while $D_{\rm BF}$ gradually increases.
In the regime of large number of fermionic atoms  (phase III),
$D_{\rm B}$ increases again and $D_{\rm F}$ also increases to a significant level.
\par
Each phase is characterized by a formation of different types of on-site composite particles\cite{Lewenstein2004} near the trap center.
The transition from one phase to another can be understood qualitatively by considering the relative strength of the various types of on-site interactions and the associated composite particles, as shown in Fig. 4.
\par
In phase I, a composite particle of two bosons and one fermion (designated type A (``BBF'')), which has the lowest on-site interaction energy, is predominantly formed 
until the number of fermions reaches the half of the number of bosons.
This phase is thus characterized as a rapid increase of $D_{\rm B}$ and $D_{\rm BF}$, 
as is observed in our experiment (Fig. 3 (b), top and bottom panels).
\par
In phase II where the number of fermionic atoms is intermediate ($N_{\rm F} > 2.5 \times 10^3$),
singly occupied fermions should appear because only a limited number of fermions can form type A particles (``BBF'').
Note that it is more energetically favorable to form two composite particles of one boson and one fermion (designated type B (``BF''))
than to form one singly occupied fermion and a type A particle (``BBF'').  
Therefore, upon an increase of the number of fermions, already formed type A particles (``BBF'') are dissolved into two type B particles (``BF'').
Namely, ``BBF''+``F'' $\rightarrow$ ``BF''+``BF''.
This results in a decrease of bosonic double occupancy and increase in boson-fermion pair occupancy, which is again consistent with our observations (Fig. 3 (b), top and bottom panels).
\par
In phase III, where the number of fermionic atoms is large, $D_{\rm F}$ begins to grow in the center due to the harmonic confinement.
It is important to note that a composite particle of two bosons and two fermions (designated type C (``BBFF'')) is energetically favorable than the case of one singly occupied boson and a composite particle of one boson and two fermions (designated type D (``BFF'')), as well as the case of two bosons, singly or doubly occupied, and one doubly occupied fermions.  
As a result, the stair-wise density distribution of fermionic atoms attracts bosonic double occupancy to the center and type C particles (``BBFF'')) are formed.
Namely, ``BFF''+``B'' $\rightarrow$ ``BBFF''.
This is again consistent with the observed increase of fermionic and also bosonic double occupancy shown in phase III of Fig. 3(b).
\par
Numerical calculations quantitatively verify the above described models. 
In fact, the density distributions of bosons ($\langle n_{\rm B, i} \rangle$) and fermions ($\langle n_{\rm F, i} \rangle$)  
(Fig. 3 (f)), 
and double occupancies of bosons ($\langle D_{\rm B, i} \rangle $) and fermions ($\langle D_{\rm B, i} \rangle$) (Fig. 3 (g)), 
and the boson-fermion pair occupancy $\langle D_{\rm BF, i} \rangle$ (Fig. 3 (h))
well reproduce the overall behaviors discussed above.
Note that the adiabatic heating, which we discuss in the next section, takes place and the boundary between each phase is rather smoothed.
\par

\section{Thermodynamics}   
As is discussed in the previous sections, the numerical calculations show reasonable agreement with the experimental data within a small uncertainty of the initial temperatures $T_{\rm ini}=40 \pm 10$ nK.
Note that the loading into the lattice is assumed to be adiabatic and thus isentropic, 
 and that the drastic change occurs in the atom temperature during this adiabatic loading process.
Here we carefully evaluate the final temperatures after lattice loading, $T_{\rm lattice}$, through the comparison between the experimental data and the numerical calculations,
and reveal the adiabatic heating and cooling effects in the attractively and repulsively interacting dual Mott insulators, respectively.
The detailed procedures of our numerical calculations are described in the Method section.
\par
In Fig. 5(a) and (b), we re-plot the measured bosonic double occupancies along with the theoretical curves for several values of $T_{\rm lattice}$ in the repulsively and attractively interacting systems, respectively.
We determine $T_{\rm lattice}$ realized in the experiments by comparing these measured $D_B$ values with calculated ones, and the results are summarized in Fig. 5(c) and (d).
We see that $T_{\rm lattice}$ ranges from 5 nK to 40 nK and from 40 nK to 100 nK for the repulsively interacting and the attractively interacting cases, respectively.
Since $T_{\rm ini}$ is roughly 40 nK in our experiments, the results in Fig. 5(c) and (d) show that cooling and heating occur in the repulsively and attractively interacting dual Mott insulators, respectively. 
Although three-body losses and  non-ideal conditions in real experiments may affect the estimated $T_{\rm lattice}$, the obtained agreement between the estimated $T_{\rm lattice}$
(filled circles in Fig. 5(c) and (d)) and calculated $T_{\rm lattice}$ under the assumption of adiabatic loading with $T_{\rm ini}$=40$\pm$10 nK(shaded area in Fig. 5(c) and (d)) suggests that the attractively (repulsively) interacting system is adiabatically heated (cooled).
\par
Here we discuss the mechanism of adiabatic cooling and heating.
In the repulsively interacting system where mixed or phase-separated Mott insulating states appear, 
the spin degrees of freedom in fermionic $^{173}$Yb can carry a large entropy of ln 6 per site in the Mott insulator.
Especially in the mixed Mott states, bosonic atoms can also contribute to the entropy as the seventh degree of freedom in addition to the fermionic atoms, resulting in a further large entropy of ln 7.
Therefore, under the isentropic condition, the adiabatic lattice loading towards the Mott regime results in the cooling of the other degrees of freedom \cite{Pomeranchuk,Werner,Cazalilla-Yb, Hazzard2010}.
Note that the state of phase-separated Mott insulators is easily destroyed by thermal fluctuations, as discussed in Supplementary Information,
and $D_B$ in the phase II rapidly decreases with the increase in $T_{\rm lattice}$ as shown by the theoretical curves in Fig. 5(a).
Thanks to the cooling effect caused by the large entropy carried by $^{173}$Yb fermions, we can successfully find the phase-separated state.
\par
On the other hand, in the attractively interacting system, the system is adiabatically heated over a wide range\cite{Cramer2008}.
As already discussed in the previous section, a key feature of the attractively interacting system is the formation of various types of composite particles,
depending on the number of fermionic atoms.
The important consequence of the formation of composite particles is the reduction of the entropy due to the reduction in the number of particles involved in the system.
Thus, the adiabaticity of the lattice loading results in the heating of the other degrees of freedom and in the melting of the Mott plateau.
We confirm the correlation between $T_{\rm lattice}$ and the fraction of the composite particles (see Supplementary Information).

		\begin{center}
		\begin{figure*}[htb]
		\includegraphics[width=15cm]{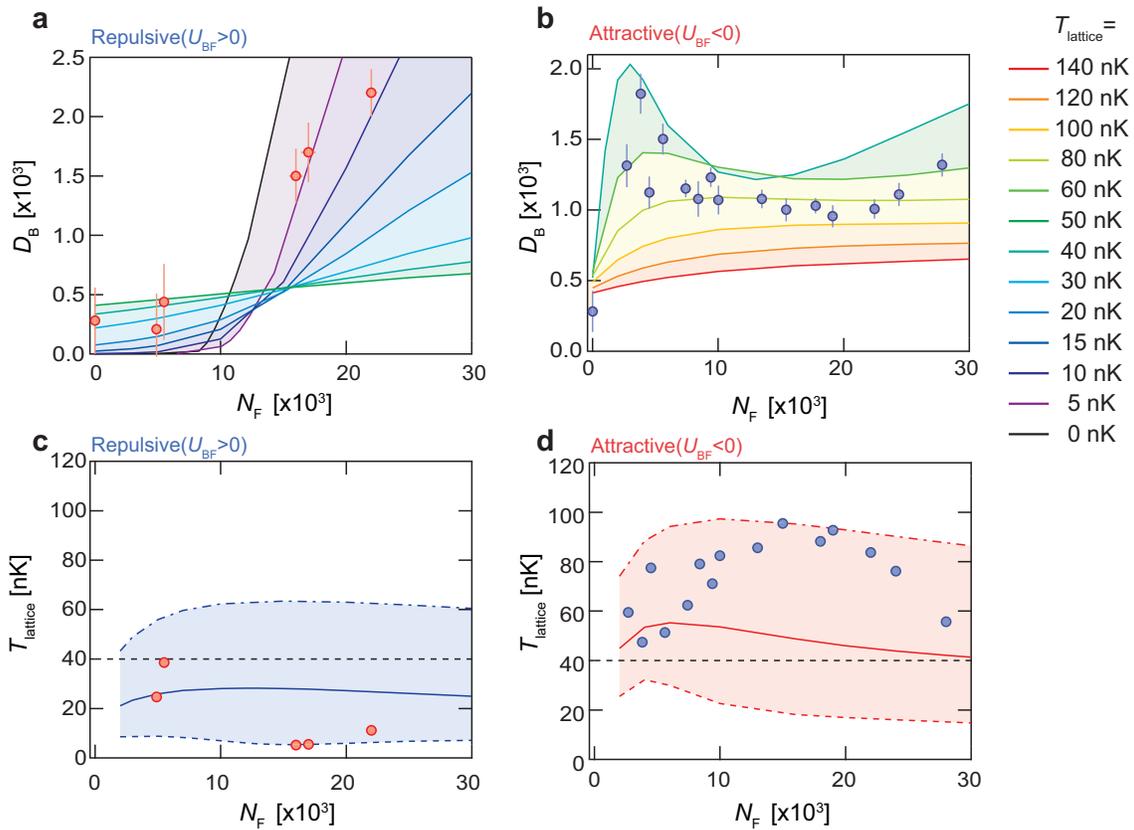}
		\caption{{\bf Thermodynamics of repulsively and attractively interacting dual Mott insulators.} 
		(a),(b) Bosonic pair occupancies for several temperatures after the lattice loading, $T_{\rm lattice}$, in
		(a) repulsively and (b) attractively interacting dual Mott insulators. The solid lines represent the calculations, and the circles are the experimental data.
		(c) $T_{\rm lattice}$ evaluated through comparing the calculations and the measurements shown in Fig. 5(a) (red circles) and calculated $T_{\rm lattice}$ for $T_{\rm ini}$=30 nK(blue dotted line), 40 nK(blue solid line), and 50 nK(blue dash-dotted line).
		(d) $T_{\rm lattice}$ evaluated through comparing the calculations and the measurements shown in Fig.5(b) (blue circles) and calculated $T_{\rm lattice}$ for $T_{\rm ini}$=30 nK(red dotted line), 40 nK(red solid line) and 50 nK(red dash-dotted line).
		}	
		\end{figure*}
		\end{center}
		
\section{Outlook}    
We have investigated quantum phases of dual Mott insulators of bosons
and fermions focusing on the roles of interspecies interactions and
relative fillings of two species.
By carefully comparing measured data with theoretical simulations, we
have successfully clarified the characteristics of several quantum
phases and the nontrivial thermodynamics during the lattice loading.
The present work paves the way for exploring diverse quantum phases
around the Mott transitions at lower temperatures
such as the alternating Mott insulator \cite{Titvinidze2008} and the
novel phases starring composite particles \cite{Lewenstein2004}.
For further investigations, we have found the important similarity
between our finding and a basic problem in condensed matter systems.
In some manganates, it is known that the metal-insulator transition
accompanies the ferromagnetism \cite{ImadaReview}.
We observe in the repulsively interacting system that bosonic atoms
undergo the crossover between the compressible and incompressible states
accompanied with phase separation.
This can be understood as a manifestation of such a ferromagnetic-type
transition, which causes the phase separation in the presence of
confinement.
This feature provides a new concept of ferromagnetism in the strongly
correlated many-body systems.

\section{Methods} 
\subsection{Preparation of dual Mott insulators of bosons and fermions with Yb isotopes}
A quantum degenerate Bose-Fermi mixture of repulsively interacting $^{174}$Yb-$^{173}$Yb or attractively interacting $^{170}$Yb-$^{173}$Yb is first created in an optical trap produced by crossed laser beams with a wavelength of 532 nm by sympathetic evaporative cooling of bosonic and fermionic species \cite{Yb_mixture}.
While we prepare the mixtures with various numbers of atoms up to 5$\times$10$^4$, the final potential depth is always set to the same value.
The final trapping frequencies of the crossed optical dipole trap are 2$\pi \times$(125, 30, 190) Hz, 
which are measured by a parametric excitation and a dipole oscillation, and the temperature is 40 $\pm$ 10 nK. 
A $^{170}$Yb single species Bose-Einstein condensate(BEC) is prepared by blasting $^{173}$Yb with a laser beam resonant to
the $^1\mathrm{S}_0\leftrightarrow {^3\mathrm{P}_1}$  transition after the production of the quantum degenerate mixture,
and subsequently waiting for thermalization for typically 100 ms to 250 ms.
Single species quantum degenerate gases of $^{174}$Yb BEC or $^{173}$Yb Fermi degeneracy are obtained by a straightforward evaporative cooling without resort to the sympathetic cooling.
After preparing the quantum degenerate mixture in the optical dipole trap, we load the atoms into a three dimensional optical lattice with a lattice constant of 266 nm in a simple cubic symmetry.
We ramp up the lattice potential smoothly to the final value within 200 ms without changing the optical dipole trap potential.
We set the final lattice depth to 15 $E_{\rm R}$ for the repulsively interacting system and 18 $E_{\rm R}$ for the attractively interacting system.  
Here, $E_{\rm R} = h^2/(2 m \lambda^2) \sim h \times 4.05$ $\mathrm{kHz}$ is the recoil energy of the lattice laser beam. 
$m$ is the mass of $^{174}$Yb and $\lambda$(= 532 nm) is the wavelength of the lattice laser.
The trap frequencies of external harmonic confinements in the optical lattice are 2$\pi\times$(145, 81, 206)Hz and 2$\pi\times$(147, 84, 209)Hz for 15E$_{\rm R}$ and 18E$_{\rm R}$, respectively.
The laser intensity of each lattice axis is precisely controlled by the feedback of  the monitored laser power.
The lattice depth is calibrated by observing the diffraction patterns due to a pulsed lattice beam.
\par
Here it is noted that one of the features of using Yb isotopes is
the negligible difference in the gravitational sags which comes from the fact that the mass and trapping potentials are almost the same for Yb isotopes.
Therefore, this clearly focuses the physics on the role of interspecies interaction between bosons and fermions
in discussing the behaviors of our dual Mott insulators. 
It is noted that on-site interaction energies between fermions and those between boson and fermion are 
independent of the fermionic spin components, realizing enlarged spin symmetry in the system \cite{Taie2010}.
In this work, we especially explore strongly-correlated regime in an optical lattice, where interaction energies dominate
the tunneling energies or thermal fluctuation. 
Due to the different s-wave scattering lengths, 
bosonic isotopes of $^{174}$Yb and $^{170}$Yb enter the Mott insulating regime of unit filling for lattice depths deeper than 11.2$E_{\rm R}$ and 12.9$E_{\rm R}$, respectively\cite{Yb_Mott, Krauth1992}.

\subsection{Double and pair occupancy measurement}   

We use one-color PA resonances associated with the intercombination transition ($^1\mathrm{S}_0+^1\mathrm{S}_0 \leftrightarrow {^1\mathrm{S}_0+^3\mathrm{P}_1}$).
The PA laser intensity and pulse duration are typically 0.5$\sim$2 W/cm$^2$ and 10$\sim$30 ms, respectively, which assures of the negligible photon scattering loss for the singly occupied sites due to the PA light.
In addition, to freeze out the site occupancy during the PA process, we ramp up the lattice depth to 25 $E_{\rm R}$
 in a short time of 2 ms which is at the same time long enough to suppress the non-adiabatic excitation to higher bands. 
The PA resonances used for the homonuclear cases of $^{170}$Yb, $^{174}$Yb, and $^{173}$Yb are located below the corresponding atomic transitions by 618 MHz, 687 MHz, and 796 MHz, respectively.
The PA resonance for the heteronuclear case of $^{174}$Yb-$^{173}$Yb ($^{170}$Yb-$^{173}$Yb)
is located below the $^1\mathrm{S}_0 \leftrightarrow {^3\mathrm{P}_1}$ of $^{174}$Yb ($^{170}$Yb) atomic transition by 674 MHz (795 MHz).
To suppress Zeeman splittings of the PA resonance due to spin degrees of freedom in $^{173}$Yb,
we reduce the residual magnetic field less than 50 mG by performing high-resolution spectroscopy associated with the magnetic sub-levels 
in the metastable ${^3\mathrm{P}_2}$ state of $^{174}$Yb.

\subsection{Single-band Hubbard model}
We determine the parameters included in the Bose-Fermi Hubbard Hamiltonian, {\it i.e.}, $U_B, U_F, U_{BF}$, and so on, based on the Bloch wave functions obtained in the {\it ab initio} manner.
From the preliminary calculations considering the higher Bloch bands, we have confirmed that the contribution of the higher Bloch bands can be negligible owing to the low filling of bosonic and fermionic atoms in the present experiments.
As a result, multiorbital effects, broadening effects of the Wannier functions \cite{Campbell2006a}, and self-trapping effects \cite{Luhmann2008}, hardly emerge in our experiments.

Some experimental parameters such as atomic mass and optical dipole trap potential take almost the same values for Yb isotopes, which significantly simplifies the model.
Thus the tunneling matrix elements $J_B$ and $J_F$ and the energy offset due to external confinement potential $\epsilon_{i}$ are set to be equal for bosons and fermions in our theoretical analysis.
We fix the single-band Hubbard parameters with the values relevant to the experiments:
$U_{\rm BB}=h\times 2.6 {\rm kHz}, U_{\rm BF}=h\times 3.5 {\rm kHz}, U_{\rm FF}=h\times 5.0 {\rm kHz}$ 
and $J=h\times 0.026 {\rm kHz}$ for the repulsively interacting system, and 
$U_{\rm BB}=h\times 2.0 {\rm kHz}, U_{\rm BF}=h\times -2.5{\rm kHz}, U_{\rm FF}=h\times 6.2 {\rm kHz}$ and $J=h\times 0.014 {\rm kHz}$ for the attractively interacting system. 

\subsection{Details of Numerical Simulation} 

We quantitatively analyze the large three-dimensional optical lattice system at finite temperatures.
Here we summarize our theoretical procedure.
We deal with the bosonic Hamiltonian based on the Gutzwiller type approximation \cite{Yamashita}:
\begin{align}
\begin{split}
\hat{H}^{\rm eff}_{\rm B} =&-\sum_{i}J_i^{\rm eff} \hat{b}_i^\dag + {\rm H.c.} 
        + \frac{U_{\rm BB}}{2} \sum_{i}  \hat{n}_{{\rm B},i} (\hat{n}_{{\rm B},i} -1) \\
				&+\sum_{i,\sigma}  \epsilon_i \hat{n}_{{\rm B},i},
\end{split}
\end{align}
where
\begin{align}
J_i^{\rm eff} &=J\sum_{j\in {\rm neighbors}}\langle \hat{b}_j \rangle.
\end{align}
In addition, since the interaction between fermionic atoms is stronger than the bosonic interaction, we neglect the hopping term of fermionic atoms.
We thus obtain the effective local Hamiltonian. 
Then, thermal states of bosons and fermions at a certain temperature $T_{\rm lattice}$ are calculated self-consistently such that the expectation values $\langle \hat{b}_j \rangle$ converges, using exact diagonalization method by appropriately eliminating the states consisting of a large number of bosons.
Here, $T_{\rm lattice}$ is the final temperature after loading atoms into the lattice.
Such a single-site approximation becomes fairly good when we focus on a deep lattice as in this work.

Currently there are no reliable methods to measure $T_{\rm lattice}$ in experiments.
We can instead evaluate this $T_{\rm lattce}$ numerically with the assumption that the lattice-loading process is adiabatic and thus isentropic \cite{Pomeranchuk, Hackermuller2010,Cramer2008,Werner}.
It is easy to calculate the entropy $S_{\rm lattice}$ of the system described by the effective Hamiltonian mentioned above by the exact diagonalization method. 
The initial entropy of trapped atomic gases without the lattice, $S_{\rm ini}$, is simply given by neglecting the interactions between trapped atoms:
\begin{align}
S_{\rm ini}/k_B=\pi^2N_F\frac{T_{\rm ini}}{T_F}+3.6N_B\left(\frac{T_{\rm
ini}}{T_{\rm c}}\right)^3,
\end{align}
where $T_{\rm ini}$ is the initial temperature of the trapped atomic gas
before lattice ramping up, $T_{\rm F}$ is the Fermi temperature, $T_{\rm
c}$ is the BEC transition temperature, and $k_B$ is the Boltzmann constant.

This approximation of neglecting atomic interactions may cause some deviations of our numerical calculations from the experimental results.
We therefore consider $T_{\rm ini}$ as a unique fitting parameter in our simulations, and introduce small uncertainty of $T_{\rm ini}$ ($\pm$10 nK) so as to include certain numerical artifacts caused by these theoretical approximations. 
In addition, it has been pointed out that, in the case of mixtures consisting of bosons and polarized fermions, the entropy of atomic gases without the lattice hardly depends on the interspecies interactions \cite{Cramer2010}.

It is useful to explain how we analyze the photoassociation (PA) loss measurements employed in the experiments.
These measurements roughly provide us with the information on bosonic, fermionic, and boson-fermion pair occupancies at each site. 
However, in the present system consisting of bosons and multi-component fermions, complex configurations of atoms are realized at finite temperatures.
Therefore, the atom loss due to PA process is given by the statistical expectation of the following local operators that count respectively the number of pairs corresponding to the bosonic and fermionic homonuclear and the heteronuclear PA measurements:
\begin{align}
\hat{D}_{B,i}&= 2 \sum_{n_{B,i},n_{F,i}} {\rm floor}(n_{B,i}/2) |n_{B,i}, n_{F,i} \rangle\langle n_{B,i}, n_{F,i}|, \\
\hat{D}_{F,i}&= 2 \sum_{n_{B,i},n_{F,i}} {\rm floor}(n_{F,i}/2) |n_{B,i}, n_{F,i} \rangle\langle n_{B,i}, n_{F,i}|, \\
\hat{D}_{BF,i}&= \sum_{n_{B,i},n_{F,i}} {\rm min}(n_{F,i},n_{B,i}) |n_{B,i}, n_{F,i} \rangle\langle n_{B,i}, n_{F,i}|, 
\end{align}
where $|n_{B,i}, n_{F,i}\rangle$ represents a local Fock state with $n_{B,i}$ bosons and $n_{F,i}$ fermions, and $n_{F,i}=\sum_\sigma n_{F,i,\sigma}$.
Here, floor($x$) is the floor function which returns the largest integer
not greater than its input, and min($x,y$) returns the minimum inputs.
We straightforwardly calculate the sums of expectation values of these local operators over the lattice sites, and compare them with the experimentally measured PA losses.
In addition to the above we further measure the heteronuclear PA loss after removing bosonic double occupancy using homonuclear PA as shown in Supplementary Information.
The local operator of this measurement is given by
\begin{align}
&\hat{D}_{B-BF,i}=\\ \notag
&\sum_{n_{B,i},n_{F,i}} {\rm min}\big(n_{F,i},
n_{B,i} \,{\rm mod}\,2\big) |n_{B,i}, n_{F,i} \rangle\langle n_{B,i}, n_{F,i}|,
\end{align}
where, mod is the modulo function.
	

\subsection{}
{\bf Acknowledgments} We acknowledge S. Uetake, T. Fukuhara, S. Sugimoto, Y. Takasu and H. Wayama for their experimental help
and J. Doyle for careful reading of the manuscript.
This work is supported by the Grant-in-Aid for Scientific Research of JSPS (No. 18204035, 21102005C01 (Quantum Cybernetics)), 
GCOE Program "The Next Generation of Physics, Spun from Universality and Emergence" from MEXT of Japan,
and World- Leading Innovative R\&D on Science and Technology (FIRST).
S.S and S.T acknowledge support from JSPS.

{\bf Additional Information} The authors declare no competing financial interests.

\clearpage
\section{Supplementary Information}

\section{Effect of thermal fluctuations in the repulsively interacting system}

In the repulsively interacting system, we find the crossover between the mixed Mott states and the phase-separated Mott states.
This can be identified by the rapid growth of  measured bosonic pair occupancy with increase in $N_F$ as shown in Fig. 3(a).
However, from the theoretical curves in Fig. 5(a), one can expect that thermal fluctuations easily blur such feature.
Our successful observation of the crossover is, therefore,  due to the cooling effect caused by the large entropy carried by fermions as mentioned in the main text.
Here we discuss in more detail how thermal fluctuations affect the crossover behavior of repulsively interacting dual Mott insulators.

\begin{center}
		\begin{figure}[bht]
		\includegraphics[width=8cm]{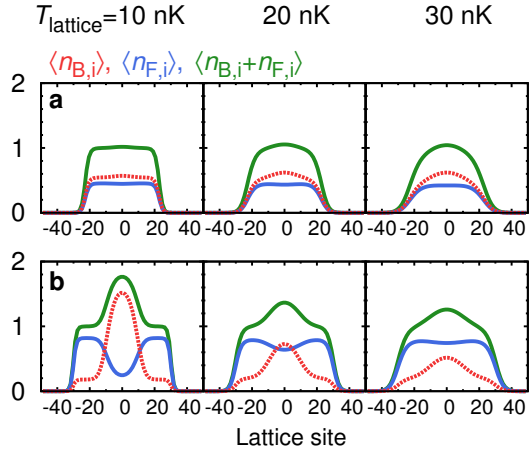}
		\caption{{\bf Temperature dependence of the spatial distribution of fermions and bosons in the repulsively interacting system. }
			(a) Mixed Mott regime with $N_{\rm B}=6\times10^3$ and $N_{\rm F}=6\times10^3$.
			(b) Phase-separated regime with $N_{\rm B}=6\times10^3$ and $N_{\rm F}=20\times10^3$.
			Temperature are chosen such that $T_{\rm lattice}=10, 20$, and $30$ nK.
		}
		\label{fig_supl_tdep}	
		\end{figure}
		\end{center}

We calculate numerically the temperature dependence of the spatial distribution of fermions and bosons. 
Figure \ref{fig_supl_tdep}(a) shows that, in the mixed Mott regime ( $N_B=6\times10^3$ and $N_F=6\times10^3$ ), the Mott plateau of total density is melted by thermal fluctuations.
This leads to the slight increase of bosonic pair occupancy with increase in temperature in this regime as confirmed from Fig. 5(a) at $N_F=6\times10^3$.
On the other hand, in the phase-separated regime ( $N_B=6\times10^3$ and $N_F=20\times10^3$ ) shown in Fig. \ref{fig_supl_tdep} (b), we find that spatial distributions of bosonic atoms become broader and broader with increase in temperature, suggesting that the phase-separated state is easily destroyed by thermal fluctuations. 
Consequently, the bosonic pair occupancy decreases rapidly with increase in temperature as shown in Fig. 5 (a) at $N_F=20\times10^3$.

		\begin{center}
		\begin{figure*}[htb]
		\includegraphics[width=6cm]{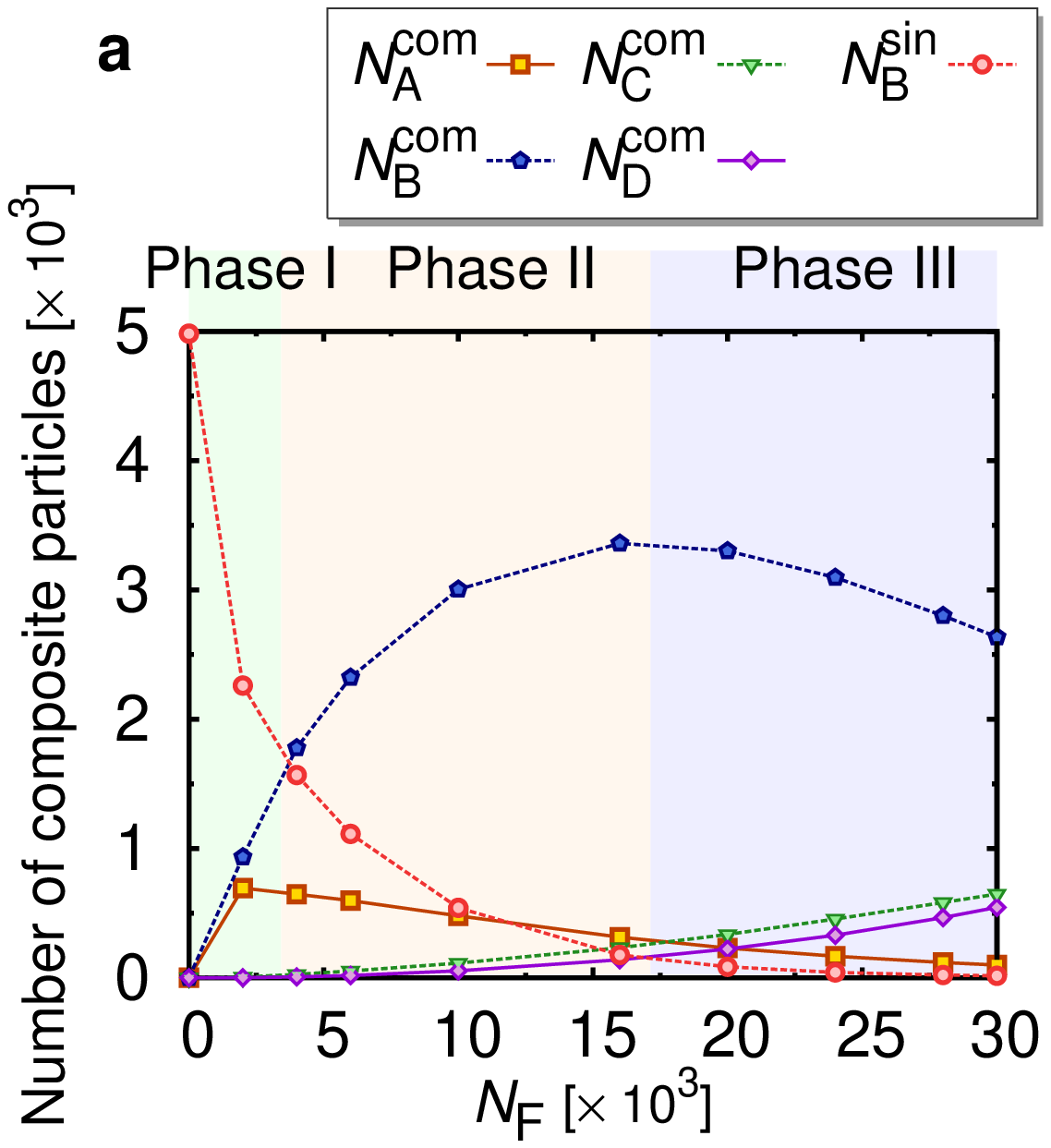}
		\includegraphics[width=9cm]{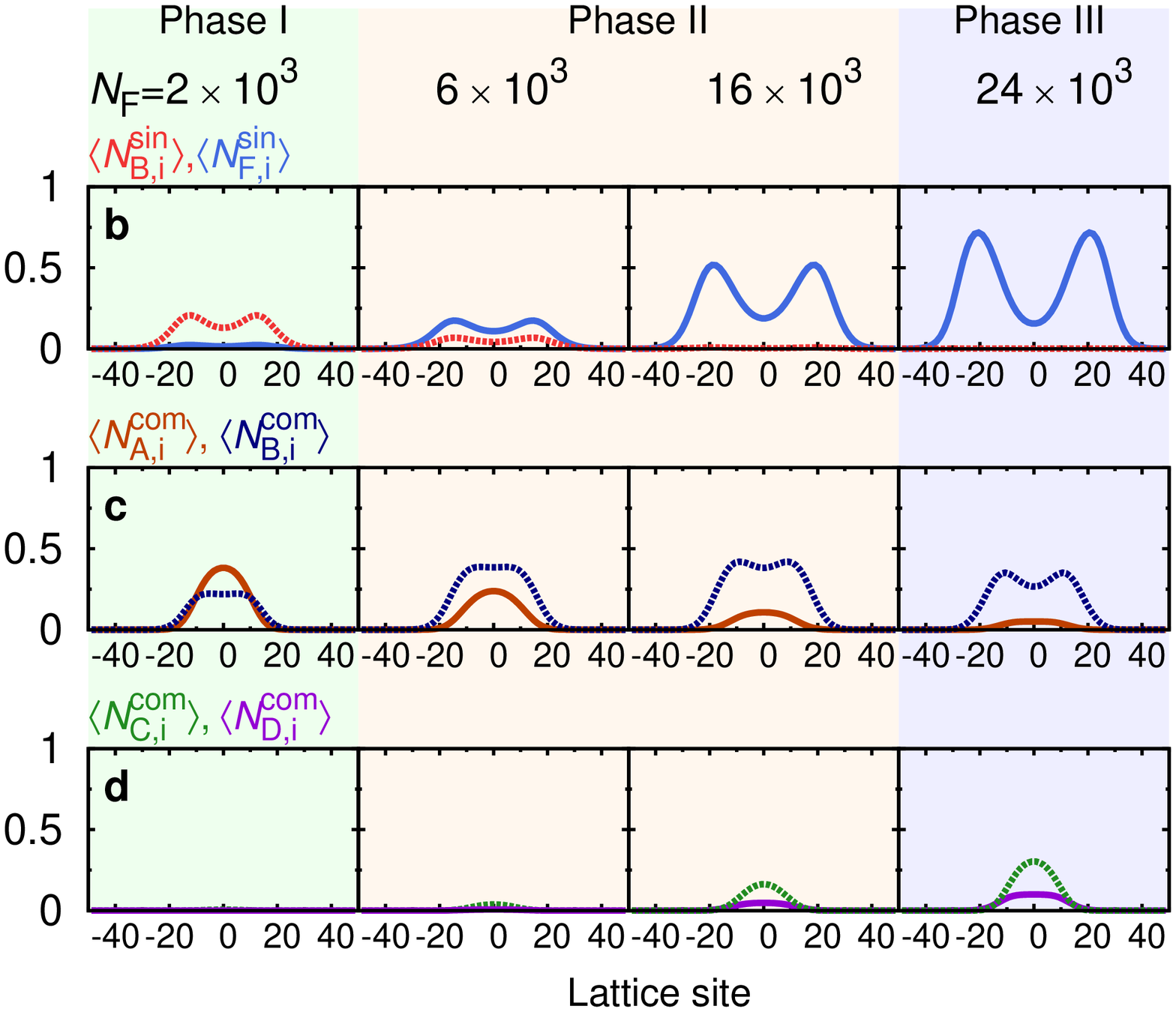}
		\caption{{\bf Distributions of the composite particles and singly occupied states. }
		(a) Number of composite particles as a function of the number of fermions $N_{\rm F}$ with $N_{\rm B}=5\times10^3$ at $T_{\rm ini}=40$ nK.
		For reference, we also plot the number of singly occupied states for bosons ($N^{\rm sin}_{\rm B}$).
		(b) Spatial distributions of the singly occupied states for bosons ($n^{\rm sin}_{B,i}$, red dashed line) and fermions ($N^{\rm sin}_{\rm F,i}$, blue line) for several $N_{\rm F}$.
        (c) Spatial distributions of the composite particles of type A(``BBF'')(($n^{\rm com}_{\rm A,i}$, orange line) and type B((``BF'')(($n^{\rm com}_{\rm B,i}$, navy dashed line).
        (d) Spatial distributions of the composite particles of type C(``BBFF'')(($n^{\rm com}_{\rm C,i}$, green dashed line) and type D((``BFF'')(($n^{\rm com}_{\rm D,i}$, purple line).
		}
		\label{fig_supl_ncp}	
		\end{figure*}
		\end{center}
		
\section{Formation of composite particles in the attractively interacting system}
In the attractively interacting dual Mott insulators, the formation of the composite particles plays a key role in the crossover of quantum phases
and also thermodynamics. 
For further understanding, it is important to see how the spatial distributions of such composite particles change for different values of $N_F$ by carefully considering the effect of the trapping potential in the experiments.
We introduce the local operators $\hat{N}^{\rm com}_{\alpha,i}$ ($\alpha=$A, B, C and D) that count the numbers of composite particles corresponding to type A(``BBF''), B(``BF''), C(``BBFF'') and D(``BFF'') defined in Fig. 4.
These operators are written as
$\hat{N}^{\rm com}_{\rm A,i}=|2, 1 \rangle\langle 2, 1|$, 
$\hat{N}^{\rm com}_{\rm B,i}=|1, 1 \rangle\langle 1, 1|$, 
$\hat{N}^{\rm com}_{\rm C,i}=|2, 2 \rangle\langle 2, 2|$, and
$\hat{N}^{\rm com}_{\rm D,i}=|1, 2 \rangle\langle 1, 2|$ (see Method for the notation).
The local number operators for singly occupied states of bosons and fermions are written as $\hat{N}^{\rm sin}_{\rm B,i}=|1, 0 \rangle\langle 1, 0|$ and $\hat{N}^{\rm sin}_{\rm F,i}=|0, 1 \rangle\langle 0,1|$, respectively.
We calculate the expectation values of these operators as a function of $N_F$ with fixed $N_{\rm B}=5\times10^3$ and $T_{\rm ini}=40$ nK.
Figure \ref{fig_supl_ncp}(a) shows the $N_F$ dependence of the total number of the type $\alpha$ composite particles, $N^{\rm com}_\alpha\equiv \sum_i\langle \hat{N}^{\rm com}_{\alpha,i}\rangle$, and singly occupied states, $N^{\rm sin}_{\beta}\equiv\sum_i\langle \hat{N}^{\rm sin}_{\beta,i}\rangle$ ($\beta=$B or F).
In Fig. \ref{fig_supl_ncp}(b)-(d), we plot the corresponding spatial distributions of singly occupied states and composite particles over the sites for several $N_F$ values.
We summarize the formation of composite particles in the three different phases discussed in the main text;

\begin{itemize}
\item
In phase I where $N_F<2.5\times 10^3$, we see from Fig. \ref{fig_supl_ncp}(a) that type A(``BBF'') and type B(``BF'') composite particles are created as $N_F$ increases.
The most energetically favorable composite particles of type A occupy the sites around the center in the trapping potential as shown in the left panel of Fig. \ref{fig_supl_ncp}(c).

\item
In phase II where the number of fermions is intermediate ($N_F>2.5\times10^3$), we see in Fig. \ref{fig_supl_ncp}(a) and (c) that  $\langle N^{\rm com}_{\rm B,i}\rangle$ and $N^{\rm com}_{\rm B}$ grow drastically with increase in $N_F$, while  $\langle N^{\rm com}_{\rm A,i}\rangle$ and $N^{\rm com}_A$ gradually decreases. 
We also find that the number of the bosonic singly occupied state ($N^{\rm sin}_{\rm B}$) decreases significantly in this phase II as shown in Fig. \ref{fig_supl_ncp}(a) and (b).
In addition, Fig. \ref{fig_supl_ncp}(a) shows that the number of composite particles becomes largest in phase II around $N_{\rm F}\sim15\times10^3$.
The similar $N_F$ dependence is observed for the final temperature $T_{\rm lattice}$ in Fig. 5(d) having a broad maximum in phase II.
This is because the total number of particles effectively decreases, and thus the entropy is reduced.

\item
In phase III where $N_F$ is large, Fig. \ref{fig_supl_ncp}(a) shows that $N^{\rm com}_{\rm B}$ gradually decreases and both $N^{\rm com}_{\rm C}$ and $N^{\rm com}_{\rm D}$ increase as $N_F$ increases.
From the right panel of Fig. \ref{fig_supl_ncp}(d), we find that the more energetically favorable composite particles of type C occupy the sites around the center in the trapping potential.
In addition, the right panel of Fig. \ref{fig_supl_ncp}(b) shows that the local density of singly occupied state of fermions $\langle N^{\rm sin}_{\rm F,i}\rangle$ grows in the outer region so as to surround the type C particles in the central region, showing a feature of phase-separation.
This originates from the effective repulsive interaction between the type C particle and the single fermion.
Due to the large entropy carried by the singly occupied fermions in the outer region, the adiabatic heating effects are partially compensated in this phase.
This results in the decrease of the realized $T_{\rm lattice}$ in the region of $N_{\rm F}>2.5\times10^4$ as shown in Fig. 5(d).
\end{itemize}

		\begin{figure}
		\begin{center}
				\includegraphics[width=8cm]{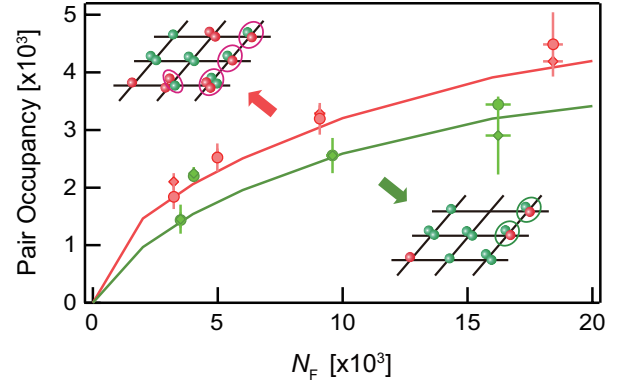}
				\caption{{\bf Measurement of contribution of composite particles of one bosons with one fermions in the attractively interacting system.}
				Boson-fermion pair occupancy (red) and boson-fermion pair occupancy after removing bosonic double occupancy (green) are measured.
				From the comparison, we can distinguish the contributions from different types of composite particles. 				
				The solid lines correspond to the numerical simulations for $T_{\rm ini}$=45 nK.
				} 
				\label{att_BB-BF}	
		\end{center}
		\end{figure}
\section{Measurement of the contributions from different composite particles}

While we have shown in Fig. 3 the boson-fermion pair occupancy measurement in the main text, 
the measurements described there cannot distinguish different types of composite particles such as type A, B, and C.
However, it is possible to discriminate the type B (``BF'') composite particle (single boson and single fermion) from other types
 which has two bosons in addition to a fermion, by performing conditional measurements.
Figure \ref{att_BB-BF} shows two kinds of boson-fermion pair occupancy measurements at the intermediate atom number of phase II,
 where the type A (``BBF'') composite particles are expected to be converted into those of type B (``BF'').
The red curve represents the measurement with the already described method.
The green one represents the boson-fermion pair occupancy measured after removing the sites with two bosons by applying the bosonic homonuclear PA laser.
The difference in the measured boson-fermion pair occupancies corresponds to the contribution from the type A (``BBF'') and type C (``BBFF'') composite particles.
The measurement clearly indicates that most of the composite particles in this regime are of type B (``BF''), which further supports our explanation. 

		\begin{center}
		\begin{figure}
		\includegraphics[width=8cm]{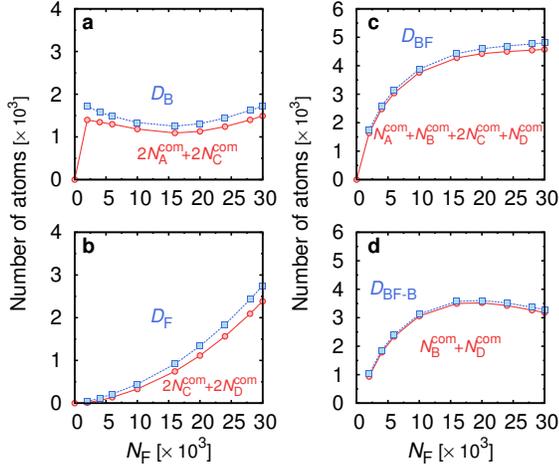}
		\caption{{\bf Number of the composite particles and pair occupancies.}
		Various calculated pair occupancies (blue) and the associated number of the composite particles (red) are plotted as a function of the number of fermions for (a) bosonic pair occupancy, (b) fermionic pair occupancy, (c) boson-fermion pair occupancy, and (d) boson-fermion pair occupancy after bosonic pairs are removed at $T_{\rm ini}=40$ nK.
		}
		\label{fig_supl_ncp_dbl}	
		\end{figure}
		\end{center}

\section{Evaluation of the composite particles numbers}
It is possible to evaluate the composite particles numbers plotted in Fig. \ref{fig_supl_ncp}(a) directly using the experimental data of PA measurements.
In our experiments, the local Fock states consisting of a large number of bosons and fermions can be reasonably neglected due to the low filling of atoms over the lattice sites.
In addition, singly occupied states of bosons and fermions are not involved in the PA losses.
We thus derive approximately the following useful relations between the pair occupancies that are measured by the PA losses and the composite particle numbers: 
\begin{align}
 D_{\rm B}&\sim  2N^{\rm com}_{\rm A}+2N^{\rm com}_{\rm C},\\
 D_{\rm F}&\sim  2N^{\rm com}_{\rm C}+2N^{\rm com}_{\rm D},\\
 D_{\rm BF}&\sim  N^{\rm com}_{\rm A}+N^{\rm com}_{\rm B}+2N^{\rm com}_{\rm C}+N^{\rm com}_{\rm D},\\
 D_{\rm BF-B}&\sim  N^{\rm com}_{\rm B}+N^{\rm com}_{\rm D}.
\end{align}
Here, $D_{\rm BF-B}$ corresponds to the boson-fermion pair occupancy obtained by the conditional PA measurement after removing bosonic pairs as explained in the previous section.
Our numerical results in Fig. \ref{fig_supl_ncp_dbl} support the validity of these relations.

\section{Pair occupancies at various lattice temperatures }

\begin{center}
		\begin{figure*}[htb]
		\includegraphics[width=14cm]{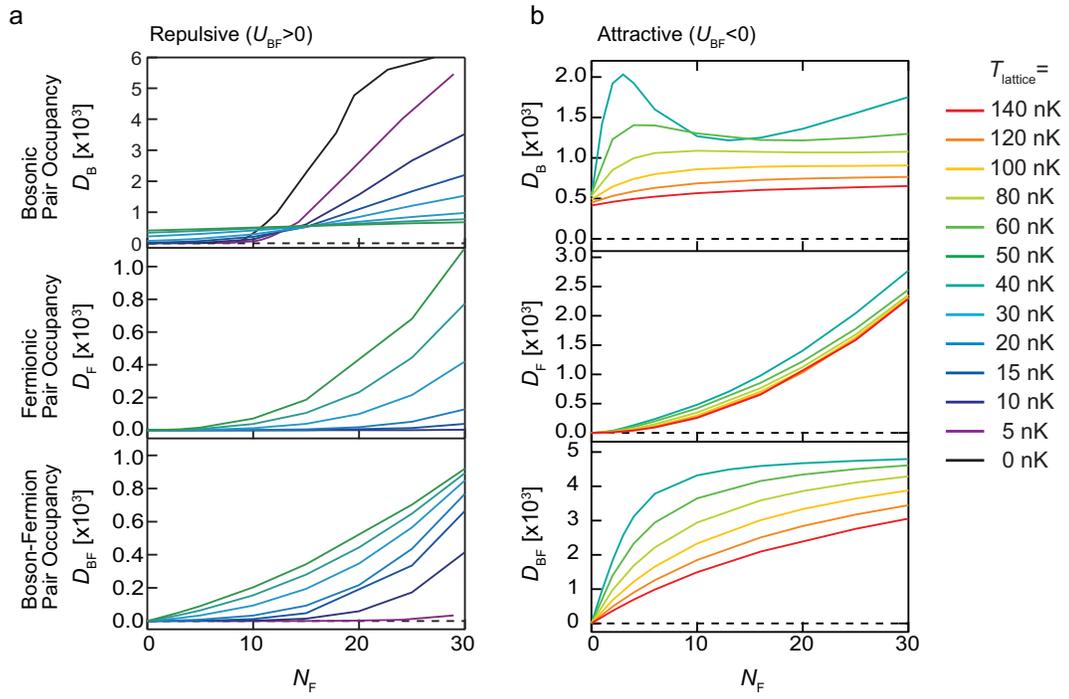}
		\caption{{\bf Numerical calculations of site-occupancies with various lattice temperatures.}}
		\label{Occupancy-Tlattice_calc}
		\end{figure*}
		\end{center}

Here we show in Fig. \ref{Occupancy-Tlattice_calc} the numerical calculations of bosonic and fermionic pair occupancies as well as boson-fermion pair occupancy with various lattice temperatures. One can also find adiabatic cooling and heating effects by comparing these numerical calculations and the experimental data shown in Fig. 3(a) and 3(b).  

\end{document}